\documentclass[twocolumn,floatfix]{aastex62}
\usepackage{gensymb}
\usepackage[super]{nth}
\usepackage[T1]{fontenc}
\usepackage[latin1]{inputenc}
\usepackage{enumerate,times,subfigure,multirow,xcolor,bm} 
\usepackage{hyperref,astrojournals,amsmath,amssymb,mathtools,graphicx,txfonts,microtype,nicefrac,xspace}

\makeatletter
\newcommand{\Md}{M_{\rm disk}}
\newcommand{\Mbh}{M_{*}}

\begin{document}
\title{Secular Eccentricity Oscillations in Axisymmetric Disks of Eccentric Orbits}

\author{Jacob Fleisig}
\affiliation{JILA and Department of Astrophysical and Planetary Sciences, CU Boulder, Boulder, CO 80309, USA}
\email{jacob.fleisig@colorado.edu}

\author{Alexander Zderic}
\affiliation{JILA and Department of Astrophysical and Planetary Sciences, CU Boulder, Boulder, CO 80309, USA}
\email{alexander.zderic@colorado.edu}

\author{Ann-Marie Madigan}
\affiliation{JILA and Department of Astrophysical and Planetary Sciences, CU Boulder, Boulder, CO 80309, USA}
\email{annmarie.madigan@colorado.edu}

\shorttitle{Eccentricity Oscillations}
\shortauthors{Fleisig, Zderic \& Madigan}
\bibliographystyle{apj} 

\begin{abstract}

Massive bodies undergo orbital eccentricity oscillations when embedded in an axisymmetric disk of smaller mass orbits. These eccentricity oscillations are driven by
secular torques that seek to equalize the apsidal precession rates of all the orbits in the disk. 
We investigate this mechanism within the context of detached objects in the outer Solar System, but we find that the oscillation timescale is too long for it to be dynamically important. It could however be interesting for phenomenon a bit farther from home; namely, feeding supermassive black holes and polluting the surfaces of white dwarf stars.

\end{abstract}



\section{INTRODUCTION}
\label{sec:intro}
Keplerian disks of eccentric orbits can undergo a rapid dynamical instability via collective gravitational forces that raises orbital inclinations, lowers eccentricities and clusters orbits in their arguments of pericenter \citep{Madigan2016,Madigan2018b}.  Minor planets in the outer Solar System show similar orbital structure \citep{Trujillo2014}, hinting at the possibility that this "inclination instability" could be responsible for dynamically reshaping their orbits. However, all simulations of the instability to date have been highly idealized. Notably, disk particles have had equal mass. Here we relax this constraint. 

In exploring the evolution of different masses undergoing the instability, we discover an additional effect: orbital eccentricity oscillations of the most massive bodies, driven by secular gravitational torques from the lower mass population. In most cases, this mechanism results in a net lowering of orbital eccentricity of massive bodies, and raising of their perihelia. We show this in section \ref{sec:results} with $N$-body simulations.
Motivated by observations of detached objects in the outer solar system \citep{Gladman2002, Trujillo2014, sheppard2016, Sheppard2019}, we refine our simulations to more closely reproduce solar system conditions.  In section \ref{sec:sims}, we simulate disks with orbits of fixed perihelia to mimic the scattered disk, we include a scattering force due to gravitational interactions with Neptune, and we draw particle masses from a continuous power-law distribution.  We find that this new dynamical mechanism is robust in all cases. 

 We discuss the physics underpinning the mechanism and its astrophysical relevance in sections \ref{sec:mech} and \ref{sec:relevance}. In particular, we derive an oscillation timescale for the new mechanism that can be applied to real systems. Under reasonable assumptions of masses of individual bodies in the outer solar system, we find that this timescale is too long to be relevant to the scattered disk.  It could however be important for Keplerian disks of larger mass, for example disks of stars around black holes or debris disks around polluted white dwarfs. 
 We conclude and discuss larger implications of this work in section \ref{sec:disc}.

\section{Simulations of eccentric disks with two masses}
\label{sec:results}

To begin, we explore the inclination instability using {\tt REBOUND} $N$-Body simulations with the IAS15 integrator \citep{Rein2012, Rein2015}.  \citet{Madigan2016} and \citet{Madigan2018b} simulated an axisymmetric disk of {\it equal mass} bodies on eccentric orbits about a massive central object. Here we improve upon these simple simulations by including bodies of different masses.

\subsection{The Inclination Instability}
\label{subsec:inc_inst}
We populate a thin ($i=10^{-4}$ radians) axisymmetric disk with $N=400$ bodies with the same orbital eccentricity, with a variety of starting eccentricities ($e = [0.50, 0.55, 0.60, 0.70, 0.80]$).  We give $399$ of them the same mass, $m$, and let just one other have mass $M=100m$.  We choose this mass ratio to create a significant difference between the low and high mass population. The mass of the disk is $\Md = 399m + M = 10^{-3} M_{*}$, where $M_{*}$ is the mass of the central object. 
 We choose a high disk mass as the instability timescale scales as $t_{\rm ins} \approx (\Mbh / \Md) ~ P$  \citep{Madigan2018b}, where $P$ denotes orbital period. 
 We place the massive body at semi-major axis $a_{M} = 1.0$, and distribute the smaller masses  uniformly between $0.9 \leq a \leq 1.1$.  $\omega$ (argument of pericenter), $\Omega$ (longitude of ascending node), and $\mathcal{M}$ (mean anomaly) are drawn from a uniform distribution $[0,2\pi)$. 
We carry out our simulations for $4000~P$, where $P$ is the orbital period of a body at semi-major axis $a=1.0$. 

\citet{Madigan2018b} showed that small-$N$ simulations inhibit secular dynamics. As we are limited here by computation time to small-$N$ simulations, our results generally overestimate secular timescales. 

\begin{figure}[!tb]
\includegraphics[width=0.43\textwidth]{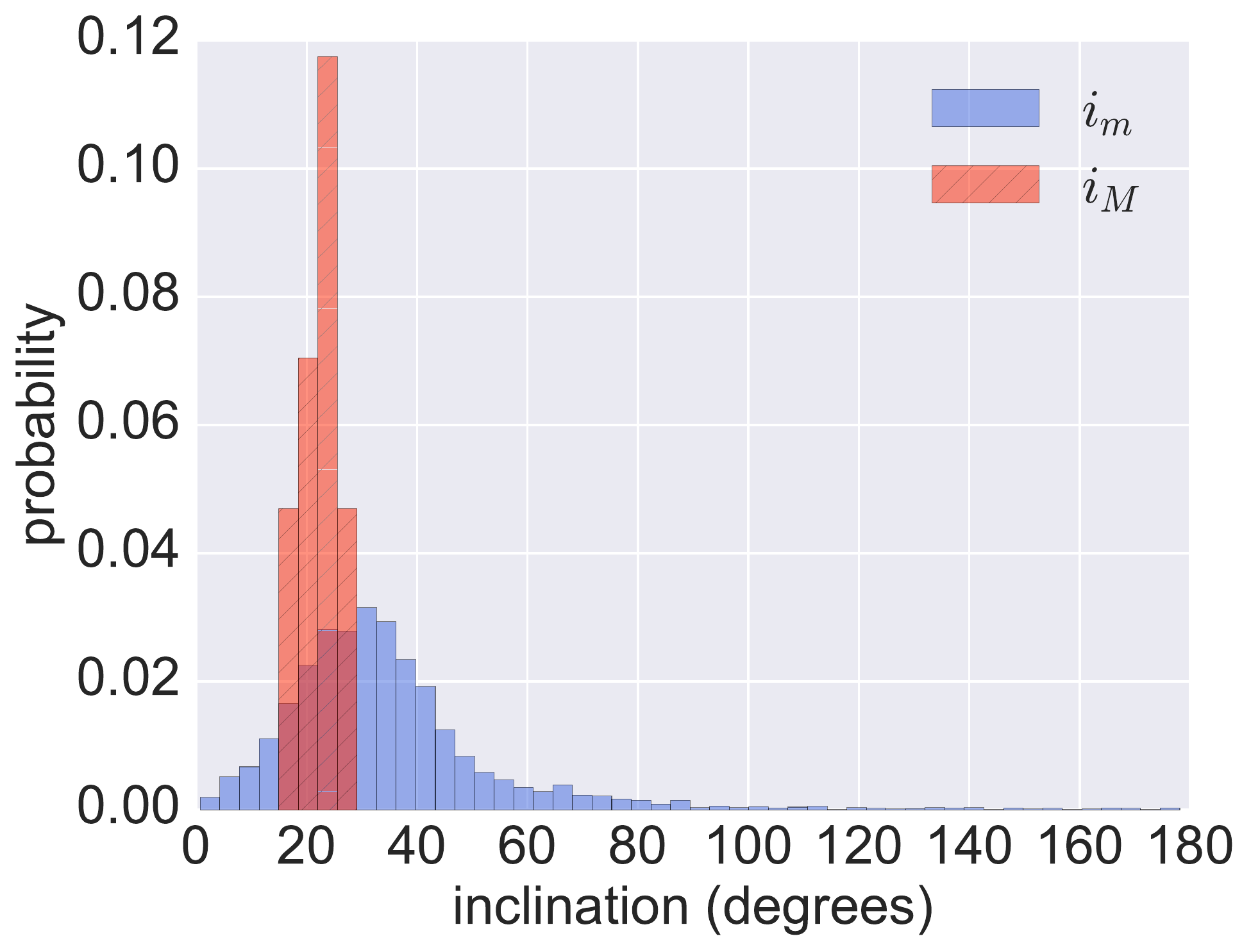}
\caption{\textbf{Inclination distributions of the large ($M$) and small ($m$) mass particles.}
The red histogram shows the probability distribution for the inclinations of the massive bodies, $i_M$, and the blue histogram shows the same for the smaller mass bodies, $i_m$. The inclination values are taken at $t \approx 3000P$, after the inclination  instability has raised the average inclination of the small mass' disk orbits to $\sim 30\degree$. These histograms combine 20 simulations with starting eccentricity $e=0.7$. The histogram of the small mass bodies contains approximately 400 times more data points than the large mass histogram (considering that each simulation with 399 small mass bodies includes only one large mass). More massive bodies tend to have lower inclinations post-instability than their less massive counterparts ($p$-value of $4.6 \times 10^{-8}$).
About $1\%$ of small mass bodies in these simulations are torqued to retrograde orbits with inclinations close to $180 \degree$ (we discuss this further in section \ref{ref:low-mass-repercussions}).
}
\vspace{0.6cm}
\label{fig:gauss}
\end{figure}

The presence of a massive body in our simulations does not inhibit the inclination instability. Interestingly, more massive bodies tend to have lower inclinations post-instability than less massive bodies. Figure \ref{fig:gauss} combines twenty simulations with one massive body with initial eccentricity $e=0.7$.  After the instability has raised the average inclination of small mass disk orbits to $\sim 30\degree$ ($t \approx 3000 P$), inclinations of the massive bodies are at lower values than that of their less massive counterparts.
We perform a Welch's t-test comparing the post-instability inclinations of the two mass populations and obtain a $p$-value of $4.56 \times 10^{-8}$. We conclude that the massive bodies' and smaller bodies' inclinations are drawn from distributions with different means.

This split in inclination distribution occurs because massive bodies drop in orbital eccentricity due to a new secular mechanism which we will explore next. 
The strength of gravitational torques decreases with eccentricity, and so the massive bodies are left with a smaller inclination after the instability has occurred. 

Thus, the inclination instability results in an inverse relationship between the mass of a body and its orbital inclination. Finding such a correlation in the population of extreme trans-Neptunian objects would strengthen the importance of collective gravitational interactions in the outer Solar System. 
We defer a discussion about observable differences in the inclination distributions of minor planets to section~\ref{subsec:detached}. 
We now discuss the new secular phenomenon that reduces the eccentricity of massive bodies. 

\begin{figure*}[tb]
\hspace{-1cm}
\includegraphics[width=1.1\textwidth]{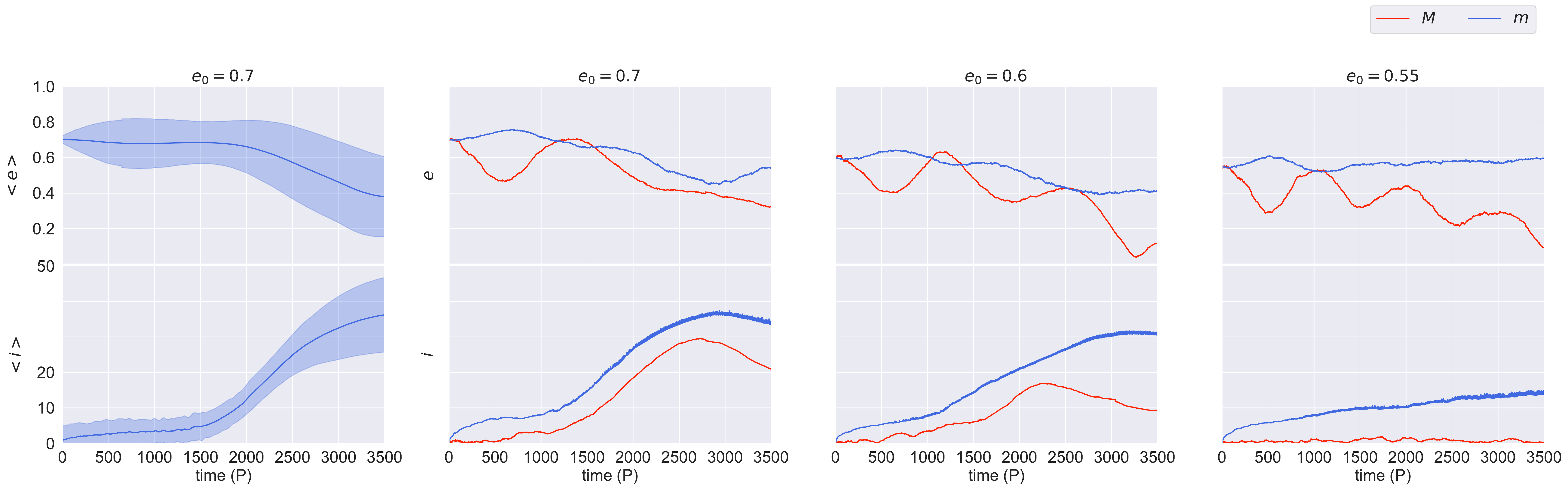}
\caption{\textbf{Eccentricity oscillations in simulations with a two-mass population.}
{Here we show the eccentricity and inclination time evolution for particles from four different simulations. 
{\it Far-left panel:} Time evolution of the mean eccentricity and inclination of particles from a simulation in which all disk particles have identical mass. Eccentricity and inclination evolve together as a result of the inclination instability. 
{\it Second-left panel:} Time evolution of the eccentricity and inclination of particles in a simulation with two mass populations.  The large mass' evolution is shown in red and the median of the small mass' evolutions are shown in blue. All particles have initial eccentricity $e_{0} = 0.7$.  Secular eccentricity oscillations emerge that are unrelated to the inclination instability. However, long term evolution of the bodies' eccentricities are still related to the inclination instability; their eccentricities decreases as their inclinations rise to conserve total angular momentum.
{\it Second-right panel:} The eccentricity and inclination time evolution of particles in a simulation with two mass populations. All particles are initiated with starting eccentricity $e_{0} = 0.6$.  Secular eccentricity oscillations unrelated to the inclination instability become more obvious.
{\it Far-right panel:} The eccentricity and inclination time evolution of particles in a simulation with two mass populations.  All particles are initiated with starting eccentricity $e_{0} = 0.55$.  When we lower the starting eccentricity of orbits in the disk further, we take the disk out of the inclination instability's unstable regime. However, the eccentricity oscillations remain.
}}
\vspace{0.6cm}
\label{fig:many_sims}
\end{figure*}

\begin{figure}[thp]
\includegraphics[width=0.45\textwidth]{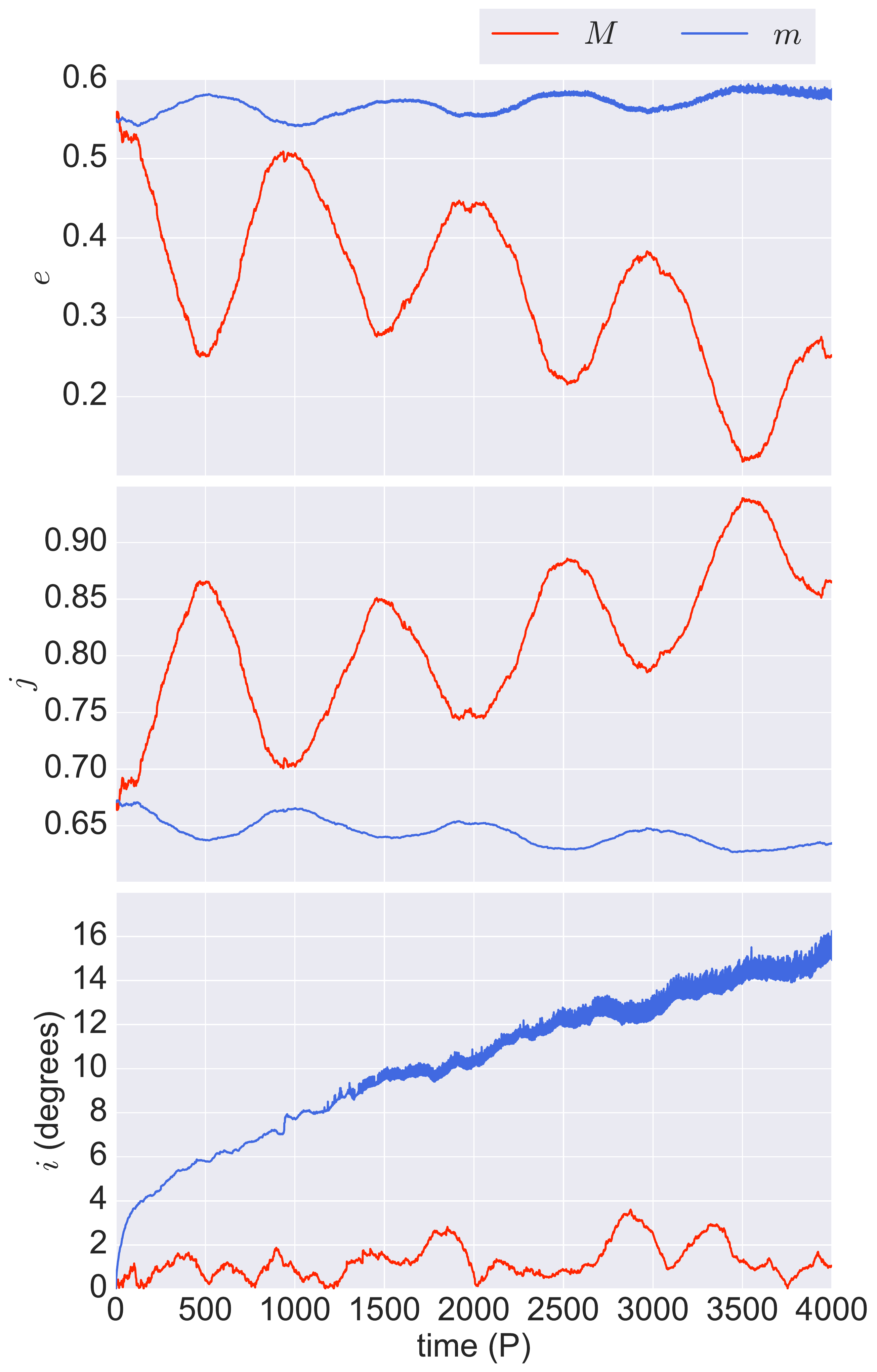}
\caption{\textbf{Secular oscillations in a simulation with initial eccentricity $\bm{e=0.55}$}.  
{\it Top panel:} The red line shows the massive body's oscillating eccentricity. The blue line shows the mean eccentricity of the smaller mass bodies oscillating in phase, but in opposite direction, with the more massive body. 
{\it Middle panel:} Specific angular momentum of the large and small mass bodies.  The massive body receives a net gain in angular momentum from the small masses.  The opposite is true for the small masses.
{\it Bottom panel:} Time evolution of orbital inclinations. We note the lack of corresponding oscillations in inclination. 
}
\label{fig:mechanism1}
\end{figure}

\subsection{Secular eccentricity oscillations}
In the far-left panel of Figure \ref{fig:many_sims}, we show the time evolution of the mean eccentricity and inclination of particles from a simulation where all disk particles have identical mass.  The eccentricity and inclination evolve together due to the inclination instability.  Orbital inclinations rise and eccentricities fall to conserve total angular momentum.  At later times ($t \sim 2600 P$) the opposite occurs, inclinations drop while eccentricities rise, as the disk relaxes to a new equilibrium state.  Eccentricity evolution in this simulation is therefore dominated by the inclination instability.

In simulations with a two-mass population, we find secular eccentricity oscillations that are not directly related to the inclination instability. 
In the second-left panel we show the eccentricity and inclination time evolution of a massive particle in a simulation with two mass populations (i.e. the simulations described in section \ref{subsec:inc_inst}) with initial eccentricity $e_{M_0} = 0.7$.  We see secular eccentricity oscillations beginning to emerge with a shorter oscillation period.  However, the long term evolution of the massive body's eccentricity is tied to the inclination instability; at $t \sim 1500 P$, its eccentricity decreases as its inclination rises.     

In order to isolate these new eccentricity oscillations, we reduce the initial eccentricities of the bodies in our simulations. This suppresses the inclination instability, since disks with $e \lesssim 0.6$ do not undergo the instability \citep{Madigan2018b}.  We show this in the second-right and far-right panels of Figure \ref{fig:many_sims} for $e_{M_0} = [0.6,0.55]$.  Here, the eccentricity evolution of the massive body is disconnected from the inclination evolution.

With the inclination  instability suppressed, we explore the oscillations further. The top panel of Figure \ref{fig:mechanism1} shows the eccentricity evolution of both the massive body and the (mean of) the small mass bodies. The mean eccentricity of the smaller bodies oscillates in phase (but in opposite direction) with the massive body's.  This is due to a secular exchange of angular momentum between the large and small mass populations (middle panel).  The large mass receives a net gain in angular momentum while the opposite is true for the small masses.  Small masses periodically donate/receive angular momentum to/from the massive body causing oscillatory behavior in both populations' eccentricities. Total angular momentum is conserved in our simulations to a factor of $10^{-5}$.

It is important to note that these eccentricity oscillations are distinct from Kozai-Lidov oscillations \citep{LIDOV1962719,1962AJ.....67..591K, Naoz2016}. The Kozai-Lidov mechanism produces orbital eccentricity and inclination oscillations of a satellite due to perturbations from a more massive, distant body. Here, eccentricity oscillations result from collective gravitational torques between orbits of similar semi-major axes, and there are no inclination oscillations due to Kozai-Lidov interactions (Figure \ref{fig:mechanism1} bottom panel).

\section{$N$-Body Simulations of Increasing Realism}
\label{sec:sims}

Motivated by observations of detached objects \--- bodies that have perihelion distances much greater than the semi-major axis of Neptune \citep{Gladman2002, Trujillo2014, sheppard2016, Sheppard2019} \--- we explore this new mechanism with respect to the outer Solar System. Could it induce eccentricity oscillations in scattered disk objects, periodically raising their perihelia and detaching them?
To test this idea, we move away from idealized simulations. We do this in three steps.

First, we initialize bodies on orbits consistent with the scattered disk population (explained below). In the second round of simulations, we add impulsive scattering events from Neptune.  In the third round, we use a distribution of masses instead of the two-mass population from our simplified set-up.

\subsection{Scattered disk profile}

We initialize particles on orbits consistent with those observed in the scattered disk.  The scattered disk, first discovered by \citet{1997Natur.387..573L}, is a collection of icy bodies that reside beyond the orbit of Neptune ($\sim 30-35$ AU to well beyond $100$ AU).  Bodies in the scattered disk are subject to strong perturbations from Neptune.  They preferentially scatter at their perihelion and, since scattered bodies come back to the location in their orbits where they received their kicks, all of them have perihelion distances on the order of Neptune's semi-major axis.  Thus, the orbital eccentricities of bodies in the scattered disk increase with increasing semi-major axes such that $p = a~(1-e) \approx \rm{constant}$.

\begin{figure}[!htb]
\includegraphics[width=0.48\textwidth]{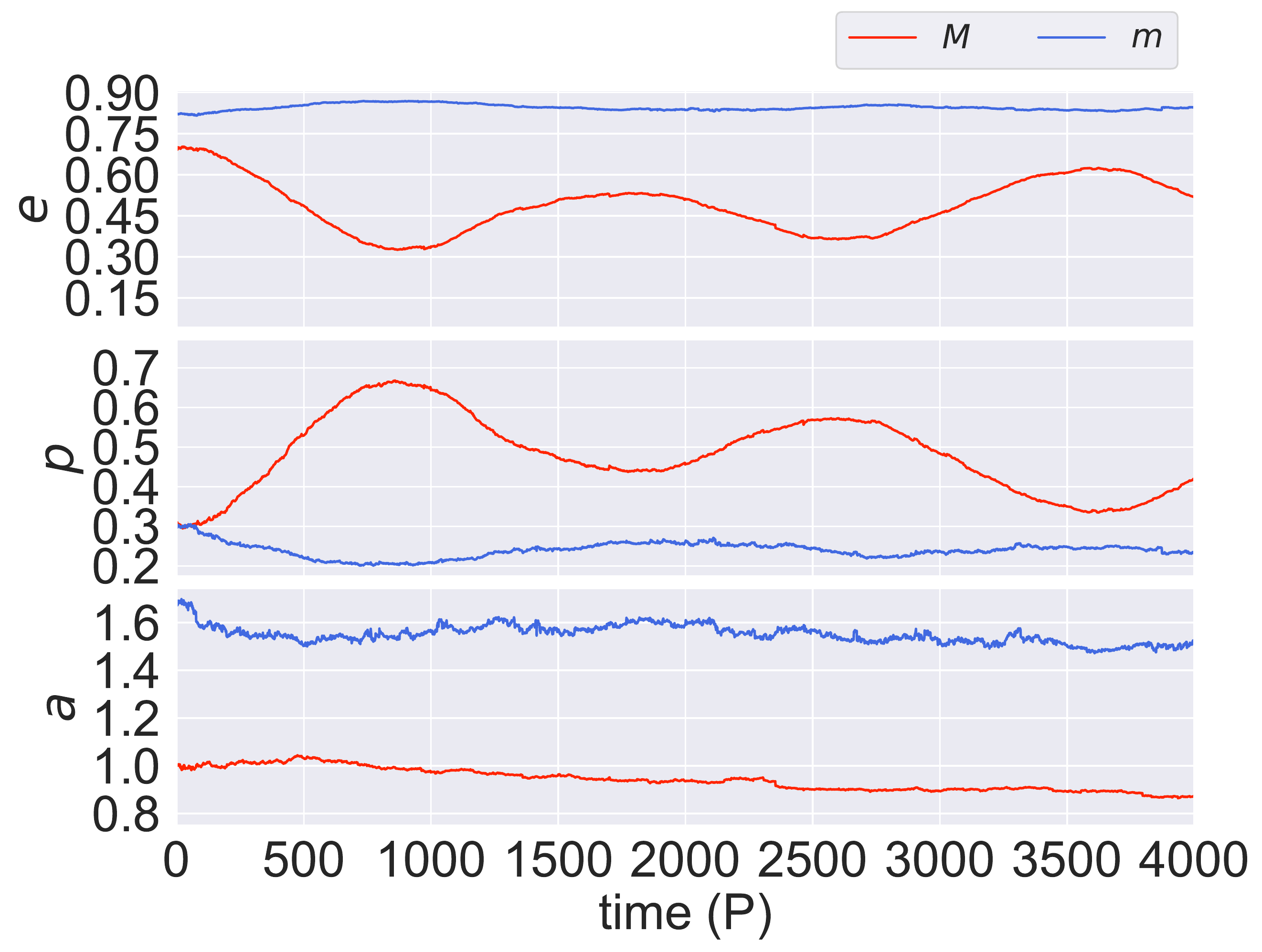}
\centering
\caption{\textbf{Eccentricity, perihelion distance, and semi-major axis evolution (scattered disk simulations).}
Eccentricity oscillations persist when the bodies are initiated on scattered disk orbits.  
When the massive body's eccentricity drops, its perihelion  distance increases.  The massive body also experiences a net drop in its semi-major axis ($a_M$) due dynamical friction.  The smaller masses exhibit opposite oscillations in eccentricity and perihelion.  The median of their semi-major axes remains roughly constant over the course of the simulations.}
\label{fig:epa_rayleigh}
\end{figure}

We reproduce this pattern in our simulations, spreading out bodies in semi-major axis space between $0.5 \leq a \leq 5.0$ according to a $1/a$ surface-density profile.  We scale our simulation parameters such that a = 1 corresponds to a = 100 AU. Our scattered bodies therefore have perihelion distances, $p = 0.3$. We assign $e$ values such that $e = 1-p/a$. We initialize inclinations with a Rayleigh distribution of mean inclination $5^\circ$.

In Figure \ref{fig:epa_rayleigh}, we show that the eccentricity oscillations persist when the bodies are initiated on scattered disk orbits.  The oscillation timescale increases due to the increased semi-major axis range of the mass distribution. The eccentricity and perihelion distance of the massive body undergo oscillations just like they do in the ideal simulations discussed in section \ref{sec:results}.  The massive body also experiences a net drop in semi-major axis ($a_M$) due to dynamical friction. The smaller masses exhibit opposite oscillations in eccentricity and perihelion.  The median of their semi-major axes remains roughly constant over the course of the simulations.

\subsection{Scattering from Neptune}
In our second round of additional simulations, we keep the initial conditions from the first but also subject the particles to scattering from a Neptune-like planet. To do this, we artificially diffuse\footnote{We do not include Neptune as an active particle in our simulations due to the high disk mass (chosen for computational efficiency).  If we assume that Neptune has a mass on the order of a million times that of a typical scattered disk object, then an active Neptune in our simulations would have mass greater than that of the central star!} the orbits' semi-major axes in our simulations using the {\tt REBOUNDX} library.  
We derive the magnitude of this artificial semi-major axis diffusion by determining the change in energy an average scattered disk object would obtain from a single interaction with a Neptune-like planet.
\citet{Duncan1987} find that, for a low inclination object with $a \approx 100$ AU, the root-mean-squared change in orbital energy is approximately $<(\Delta x)^2>^{1/2} \approx 2 \times 10^{-5}$ AU\textsuperscript{-1}.  The resulting scattering diffusion timescale is $t \approx 2.5 \times 10^5~P$. The fractional change in $a$ per orbital period is therefore $<\Delta a>/a \approx (\Delta t /t)^{1/2}$ which is equivalent to $2 \times 10^{-3}$.

We subject all the bodies in our simulations to this fractional change in $a$ every perihelion passage if their perihelion distances are $p \leq 0.35$.  In the context of the Solar System, this means we scatter our particles if their perihelion distances pass within $\sim 5$ AU of Neptune's orbit.
We choose this limit to model strong scattering from Neptune.  However, this does not model the weaker scattering events which can lead to diffusion in the semi-major axes of bodies with $p>35$ AU \citep{Bannister2017}.
While supplying the kicks in semi-major axis, we change their eccentricities such that their perihelia are fixed.  There is an equal probability that the bodies are scattered inward or outward. 

\begin{figure}[t!] 
\includegraphics[width=0.49\textwidth]{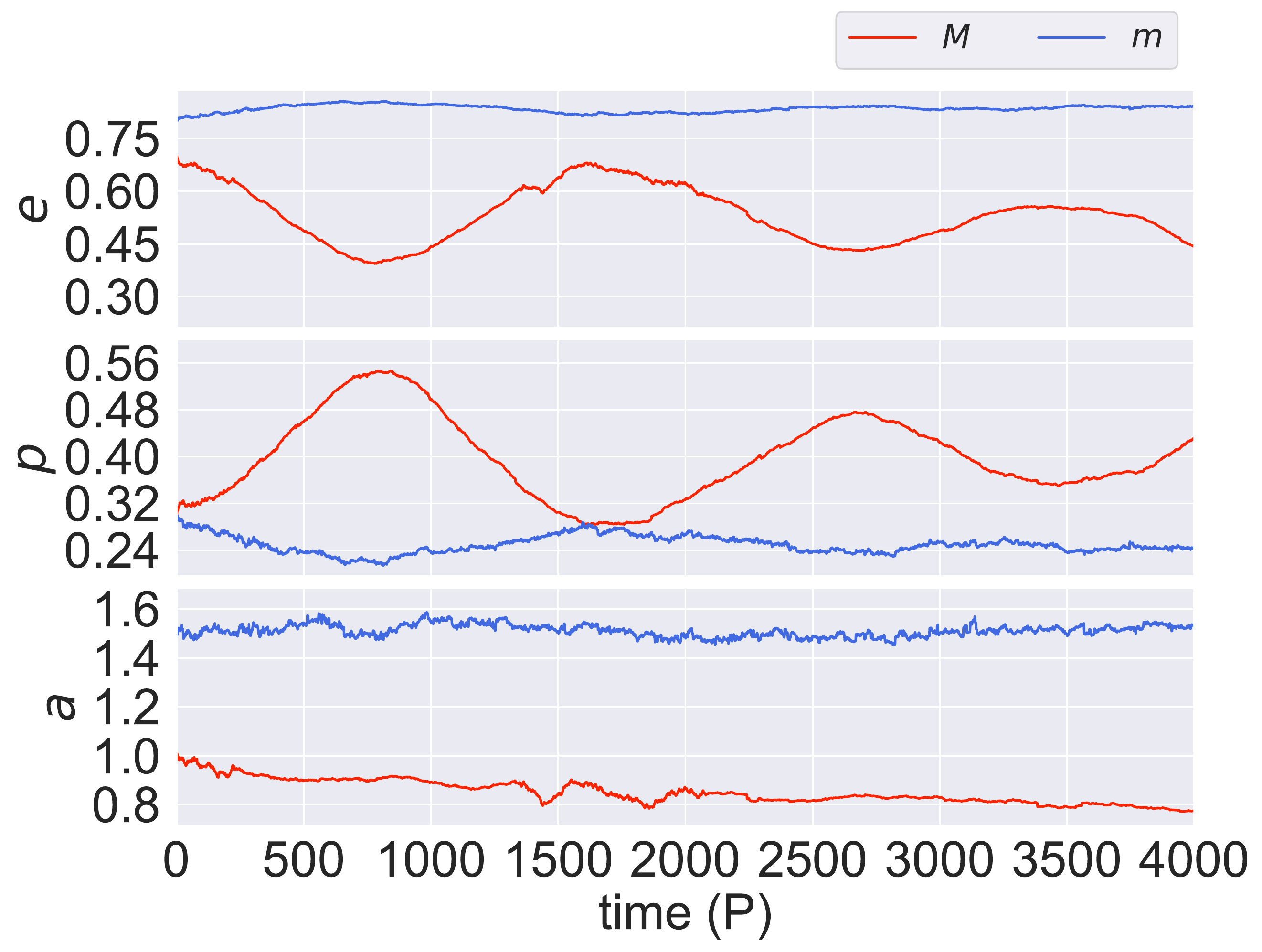}
\caption{\textbf{Eccentricity, perihelion distance, and semi-major axis evolution of the massive body (scattering simulations).}
Eccentricity oscillations persist even in the presence of continual scattering from a Neptune-like planet.  Bodies are given a kick every perihelion passage if their perihelion distance drops to $p \leq 0.35$. The massive body also experiences a net drop in its semi-major axis ($a_M$) due to dynamical friction.  The small mass bodies exhibit opposite oscillations in eccentricity and perihelion.  The median of their semi-major axes remains roughly constant over the course of the simulations.}
\label{fig:epaneptune}
\end{figure}

In Figure \ref{fig:epaneptune}, we show that this scattering force does not greatly affect the detachment mechanism.  The massive body continues to experience eccentricity and perihelion oscillations even with continual kicks in semi-major axis.   
The fact that the new secular mechanism is still robust is not so surprising. The scattering diffusion timescale in our simulations is $2.5 \times 10^5$ orbital periods.  This is the length of time required for a body to experience a change in semi-major axis proportional to its initial $a$ value. Since the secular torques in our simulations change orbits on timescales of $\sim 10^3$ orbital periods, scattering does not change the bodies' semi-major axes rapidly enough to effect the secular mechanism. Scattering also does not significantly change the orientation of the orbits (both in our simulations, and in reality), allowing the secular torques to keep acting in the same direction. 

It is important to note that this result depends on the disk mass.  Here, we simulate a scattered disk which is $\sim 10^3$ times more massive than the real scattered disk. For lower mass disks, significant diffusion through scattering could occur before secular effects have time to reshape the bodies' orbits.

\subsection{Distribution of masses}
\label{subsec:mass_distribution}
 In our third round of simulations, we initialize bodies on scattered disk orbits and assign masses according to a power-law distribution. There is no scattering force in these simulations.   
 
 Using observations and collisional models to probe the histories of Kuiper Belt objects, \citet{1538-3881-128-4-1916} estimate that the number density of Kuiper Belt objects scales as $dN/dr \propto r^{-\beta}$ where $\beta \approx 2.5-3.0$ for small bodies (with radii $r\lesssim 0.1-1.0$ km) and $\beta \approx 3.5$ for large bodies (with radii $r\gtrsim 10-100$ km).  Our aim is not to emulate this exact broken power law in our simulations but to show that this mechanism is robust for a realistic mass distribution.  If we take an intermediate value of $\beta=3.0$ and hold the density of the bodies constant, then $dN/dm \propto m^{-5/3}$. We randomly assign mass values from this density profile.

\begin{figure}[b!]
\includegraphics[width=0.47\textwidth]{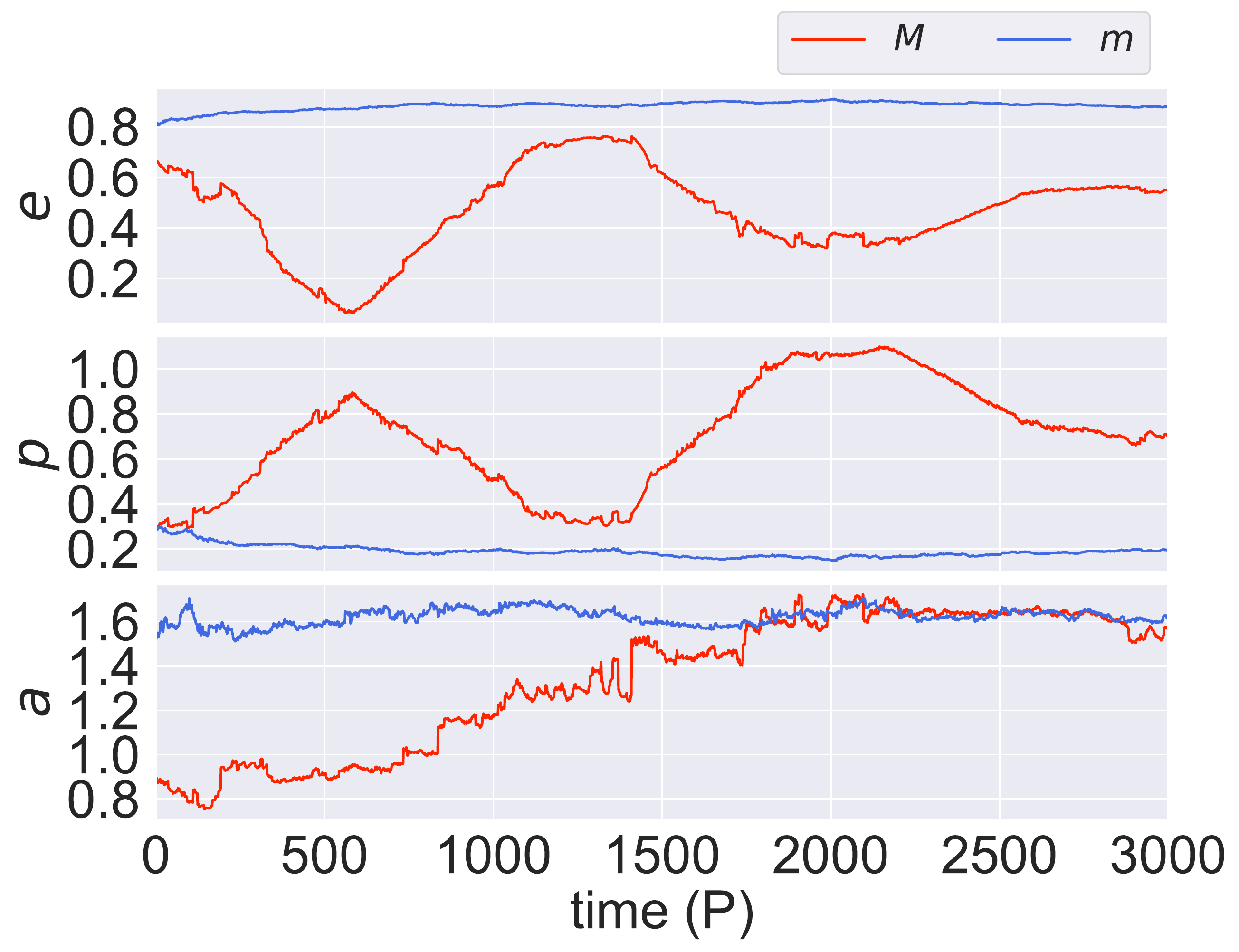}
\caption{\textbf{Eccentricity, perihelion distance, and semi-major axis of the massive body vs. time (mass distribution simulations).} 
Eccentricity oscillations persist for a realistic mass distribution.  Eccentricity and perihelion of the massive body undergo downward and upward trends respectively while the opposite occurs for smaller masses.  The semi-major axis of the largest mass evolves to larger values with time while the median semi-mjaor axis of the smaller masses remains roughly constant.  We assign masses in accordance with a realistic power law size distribution where $dN/dr \propto r^{-3}$ and $dN/dm \propto m^{-5/3}$.}
\label{fig:epa_mass_dist}
\end{figure}

In Figure \ref{fig:epa_mass_dist} we show that the mechanism is robust with this mass distribution. The most massive body in these simulations (with mass $M \approx 3 \times 10^{-4} \Mbh$ such that $M/\Md \approx 0.3$) undergoes the same cyclical orbital changes in eccentricity and perihelion. In contrast to previous simulations, the semi-major axis of the largest body evolves to larger values with time.  In fact, the massive body's semi-major axis evolves to the median value of that of the smaller masses.  The semi-major axes of the smaller mass bodies, on average, remains constant over the course of the simulations.

\section{PHYSICS BEHIND OSCILLATIONS}
\label{sec:mech}

In this section, we describe the physics behind the eccentricity oscillations.  First, we give a qualitative overview using the idealized simulations described in section \ref{sec:intro}.  We then derive analytic expressions governing the oscillations.  Finally, we discuss how this mechanism effects the small mass bodies.

\subsection{General Description}
\begin{figure}[!t]
\includegraphics[width=0.45\textwidth]{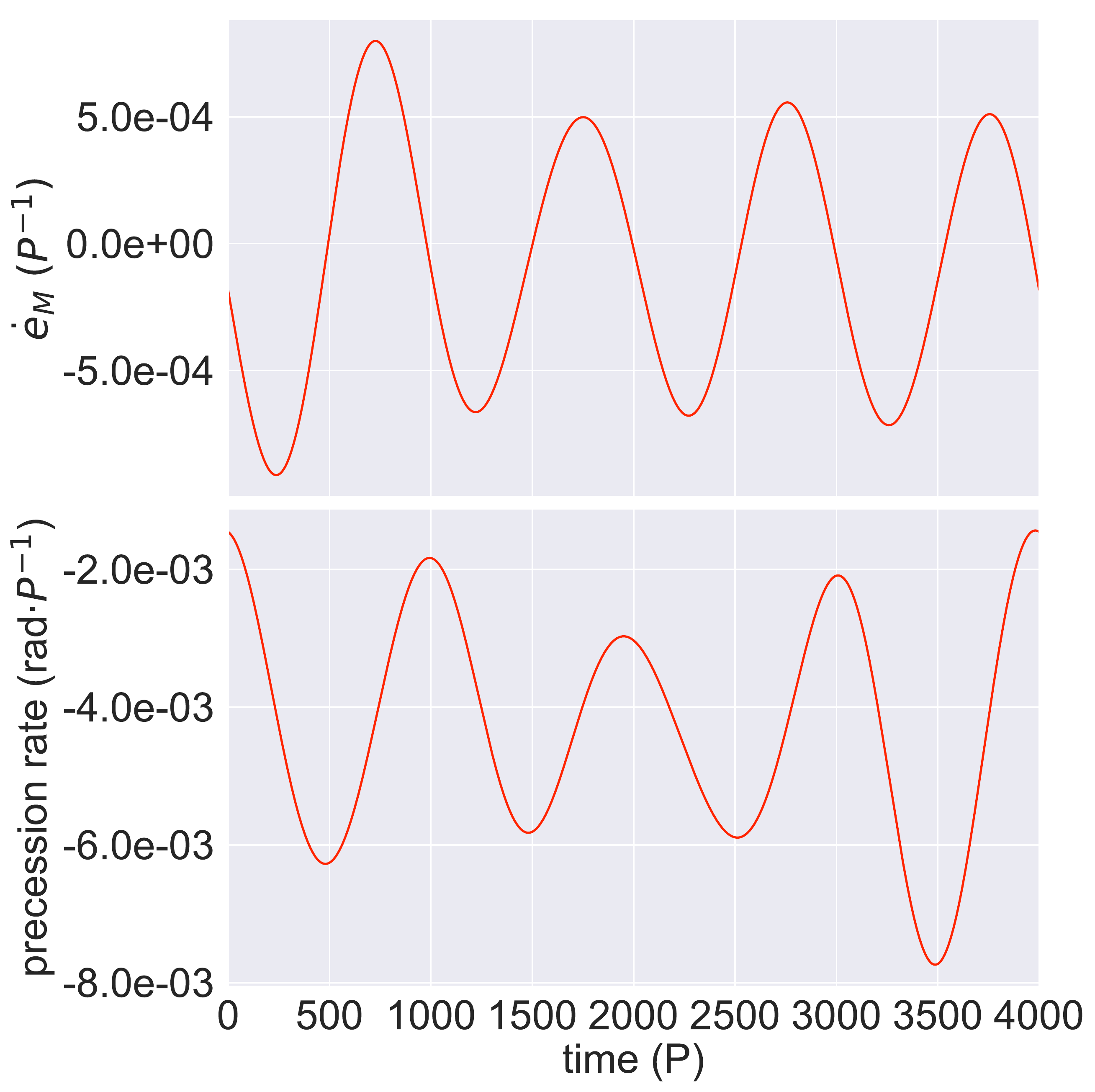}
\caption{\textbf{Rate of Change of Eccentricity and Apsidal Precession (\bm{$a=[0.9,1.1]$}, \bm{$e=0.55$}).} 
\textit{Top panel:} The massive body's rate of change of eccentricity as a function of time.
\textit{Bottom panel:}  The massive body's apsidal precession rate as a function of time.  
The massive body, initially precessing more slowly than the rest of the disk orbits, experiences secular torques from an over-density of smaller orbits on one side of {its} orbit. It gains angular momentum, its eccentricity drops and its orbit begins to precess more quickly.  The massive body, now precessing faster, gives angular momentum back to the sea of low mass particles.  Consequently, its precession rate drops and its eccentricity increases.  This secular exchange of angular momentum causes periodic oscillations in the massive body's apsidal precession rate and eccentricity.}
\label{fig:mechanism_fig2}
\end{figure}

Figure \ref{fig:mechanism_fig2} illustrates qualitatively how this mechanism operates. This plot is drawn from the simulations described in section \ref{sec:results} with $a=[0.9,1.1]$ and $e=0.55$.  
All the orbits undergo apsidal precession, due to the gravitational forces from other particles, such that their eccentricity vectors rotate with retrograde motion.  
We calculate the massive body's apsidal precession rate by first obtaining the projection of its eccentricity vector in the $xy$-plane, $i_e = \arctan ({e_y},{e_x})$ \citep{Madigan2016}. We calculate the numerical derivative of $i_e$ and perform a discrete Fourier transform, subtracting out high frequency noise components. This is shown in the  bottom panel of Figure \ref{fig:mechanism_fig2}.  We follow a similar procedure to obtain the derivative of the massive object's eccentricity (Figure \ref{fig:mechanism_fig2} top panel).

Initially, the massive body's orbit precesses slowly relative to the other bodies.  This is because the net non-Keplerian force it experiences from the small masses is smaller than the net non-Keplerian force experienced by the small masses.  Because of it's relatively slow precession, it encounters an over-density of smaller mass' orbits behind {its} orbit.  These orbits negatively torque the massive body's orbit, reducing its orbital eccentricity (Figure \ref{fig:mechanism_fig2}, top panel) and causing it to precess more quickly (due to its smaller eccentricity vector; Figure \ref{fig:mechanism_fig2} bottom panel).  After 500 orbital periods, the massive body, which is now precessing more rapidly, catches up to the smaller mass' orbits and donates angular momentum back to them.  Secular torques now act in the opposite direction as before and the massive body increases its orbital eccentricity. 
A useful way to think about this mechanism is that the secular angular momentum exchange between orbits seeks to equalize the apsidal precession rates of all the particles in the disk. Without a frictional force to damp them however, angular momentum (and hence eccentricity) oscillations persist. 

In (most of) our simulations, the massive body also experiences a net drop in eccentricity over the course of thousands of orbits.  This is due to dynamical friction.  As the massive body sinks towards the center of the gravitational potential (via two body interactions), there is less mass present within its orbit. Hence, it precesses even more slowly than before. The nearby smaller mass orbits precess faster however, due to its presence. Therefore, the relative precession rate between the massive and less massive bodies increases; the massive body must plunge to increasingly lower eccentricities to precess at the same rate. This causes its eccentricity to oscillate around increasingly smaller values.
We note that the massive body's eccentricity oscillations are centered around a value that is always lower than its initial eccentricity (if all bodies start at the same initial eccentricity). Hence the massive body tends to detach, even without dynamical friction.

\subsection{Oscillation Timescale}
In this section, we derive approximate analytic formulae that describe the new mechanism.  We again consider the idealized set-up of an axisymmetric disk consisting of a large number ($n$) of small mass bodies ($m$) on high eccentricity ($e_{m_0} = e_{m}[t = 0]$) orbits about a central mass ($\Mbh >> m$). Into this system, we place a single body of intermediate mass ($M > m$) with initial orbital eccentricity ($e_{M_0}$). For simplicity, we assume all bodies have the same semi-major axis and orbital plane.

The orbits will apsidally precess due to their mutual gravitational forces. If orbit $M$ precesses at a different rate with respect to the lower mass orbits, it will experience a flow of orbits to one side and exchange angular momentum with them via secular torques. Its precession rate over one orbital period is 
\begin{equation}
\label{eq:prec-rate-M}
\Gamma_M \approx \frac{M_{\rm disk}(<a)}{\Mbh}\sqrt{1 - e_{M_0}^2} =  \frac{n m}{\Mbh}\sqrt{1 - e_{M_0}^2}
\end{equation}
\citep{2010PhRvD..81f2002M}, where $M_{\rm disk}(<a)$ is the mass of the disk within its orbit, while the precession rate of small mass bodies is
\begin{equation}
\label{eq:prec-rate-m}
\Gamma_m \approx  \frac{n m + M}{\Mbh}\sqrt{1 - e_{m_0}^2}.
\end{equation}
We note that the equations for precession rate are approximate as they are derived assuming a spherical gravitational potential.
A better expression for the precession rate can be found for the disk of small mass bodies in \citet{Kondratyev2014}. However, we find that the eccentricity dependence of the precession rate remains unchanged and the calculation of the equilibrium eccentricity in eq.~\ref{eq:e_M} is the same. This is because we treat the massive body's contribution to the precession rate in a manner identical to the disk of small bodies. A more accurate analysis would require a numerical approach. 

If $e_{M_0} = e_{m_0}$, orbit $M$ will precess at a slower rate than orbits $m$.  To precess at the same rate, it will need to decrease its orbital eccentricity to a certain equilibrium value (which it will do via the exchange of angular momentum with surrounding orbits).
We calculate this equilibrium eccentricity by setting $\Gamma_{\rm diff} = \Gamma_m - \Gamma_M = 0$, which yields
\begin{equation}
\label{eq:e_M}
\sqrt{ 1 - e_{\rm eq_M}^2} = \left(\frac{nm + M}{nm}\right) \sqrt{1 - e_{m_t}^2}.
\end{equation}
$e_{\rm eq_M}$ is the eccentricity at which the massive orbit $M$ would precess at the same rate as the lower mass orbits.  Note however that the eccentricity of lower mass bodies at time $t$, $e_{m_t} \ne e_{m_0}$. As total angular momentum of the system is conserved, the more massive body cannot reach the equilibrium orbital eccentricity without changing the average eccentricity of the surrounding smaller mass orbits.  Conservation of orbital angular momentum yields 
\begin{equation}
\label{eq:J-cons}
J_M + nJ_m = M J_c \sqrt{1 - e_{M_t}^2}  + n m J_c  \sqrt{1 - e_{m_t}^2}  = J_{\rm tot}, 
\end{equation}
where $J_c = \sqrt{G \Mbh a}$ is the specific circular angular momentum and $J_{\rm tot}$ is the conserved total angular momentum. 
Combining equations \ref{eq:e_M} and \ref{eq:J-cons} and taking $e_{M_0} = e_{m_0}$ (appropriate for our particular simulations) yields the expression for equilibrium eccentricity
\begin{equation}
\label{eq:e_eq}
\sqrt{ 1 - e_{\rm eq_M}^2} = \delta ~ \sqrt{ 1 - e_{\rm m_0}^2},
\end{equation}
where $\delta$ is an interaction term between the two populations,
\begin{equation}
\label{eq:delta}
\delta = \frac{(1 + \alpha)^2}{\alpha^2 + (1 + \alpha)},
\end{equation}
and
\begin{equation}
\label{eq:alpha}
\alpha = \frac{nm}{M}.
\end{equation}

We note that equation~\ref{eq:e_eq} generally under-predicts the equilibrium eccentricity for a massive body as compared to our simulations. This is due to the approximation we use for the apsidal precession rate in equations \ref{eq:prec-rate-M} and \ref{eq:prec-rate-m}. As we change the initial orbital eccentricity of the massive body in a simulation, we also change the initial orbital eccentricities of the smaller mass bodies. This alters the gravitational potential which affects precession rates but is not accounted for in these equations.  

Equation \ref{eq:e_eq} predicts imaginary values for the equilibrium eccentricity of the massive body for disks with initial eccentricities below $e \sim 0.64$. This is at odds with our numerical simulation results where $e=[0.6,0.55]$ (Figure \ref{fig:many_sims}).  The massive body in these simulations still undergoes eccentricity oscillations about \textit{real} equilibrium values.  Though off by a factor of $\lesssim \!2$, the imaginary values predicted by equation \ref{eq:e_eq} points to interesting physical behavior. Massive bodies with near-circular orbits cannot drop their eccentricities low enough to precess with the rest of the disk (since their orbits cannot become more circular than a perfect circle).  In this case, their equilibrium eccentricity values would in fact be imaginary.

We calculate the oscillation period by quantifying the time it takes the massive orbit to reach the equilibrium eccentricity via the exchange of angular momentum with surrounding lower mass orbits; this is a quarter of the oscillation period.  
The change in specific angular momentum due to torques exerted on orbit $M$ over one orbital period is given by 
\begin{equation}
\Delta J_p \approx 2 \pi ~ \chi ~ \beta(e) ~\frac{n m}{\Mbh} ~ J_c,
\label{eq:jp}
\end{equation}
where $\chi ~ (< 0.5)$ is the fraction of disk orbits that donates a net positive angular momentum to orbit $M$, and $\beta(e) \approx 0.25e$ is the eccentricity dependent factor that influences the strength of the secular torque \citep{2007MNRAS.379.1083G}. To derive equation~\ref{eq:jp}, we use an expression for torque that ignores the azimuthal dependence. 
From the idealized simulations described in section \ref{sec:results}, we find that $\chi \approx 0.3$. The oscillation period is then
\begin{equation}
\label{eq:p-osc}
t_{\rm osc} = \frac{8}{\pi e_{\rm m_0}} ~  \frac{\Delta J}{J_{c}} ~ \frac{\Mbh}{\chi n m} P,
\end{equation}
where $\Delta J = J_c ~ \sqrt{1-e_{\rm eq_M}^2}- J_c ~ \sqrt{1-e_{M_0}^2}$ is the  change in specific angular momentum needed to reach the equilibrium eccentricity.  The oscillation timescale for our simulations with $e_{m_0} = e_{M_0} = 0.7$ and a corresponding equilibrium eccentricity of $e_{eq} = 0.31$ should be $\sim 3 \times 10^3 P$, which is in rough agreement with our numerical results.  However, if we use the equilibrium eccentricity from the simulation results instead of equation \ref{eq:e_eq} (i.e. $e_{eq} \approx 0.55$), then the oscillation timescale becomes $t_{osc} \sim 1.5 \times 10^3 P$; which is in much better agreement with our results (see top second-left panel of Figure \ref{fig:many_sims}). Note from Figure \ref{fig:many_sims} that $e_{eq} \approx 0.55$ is not the lowest eccentricity value the massive body obtains.  However, it is the equilibrium eccentricity associated with a fixed semi-major axis at $a=1$.  Since the massive body evolves to lower $a$ via dynamical friction, its equilibrium eccentricity changes.  In this analysis, we assume a fixed $a$ for a massive body.


\subsection{Consequences for small mass bodies}
\label{ref:low-mass-repercussions}

\begin{figure}[tp]
\centering
\includegraphics[width=0.48\textwidth]{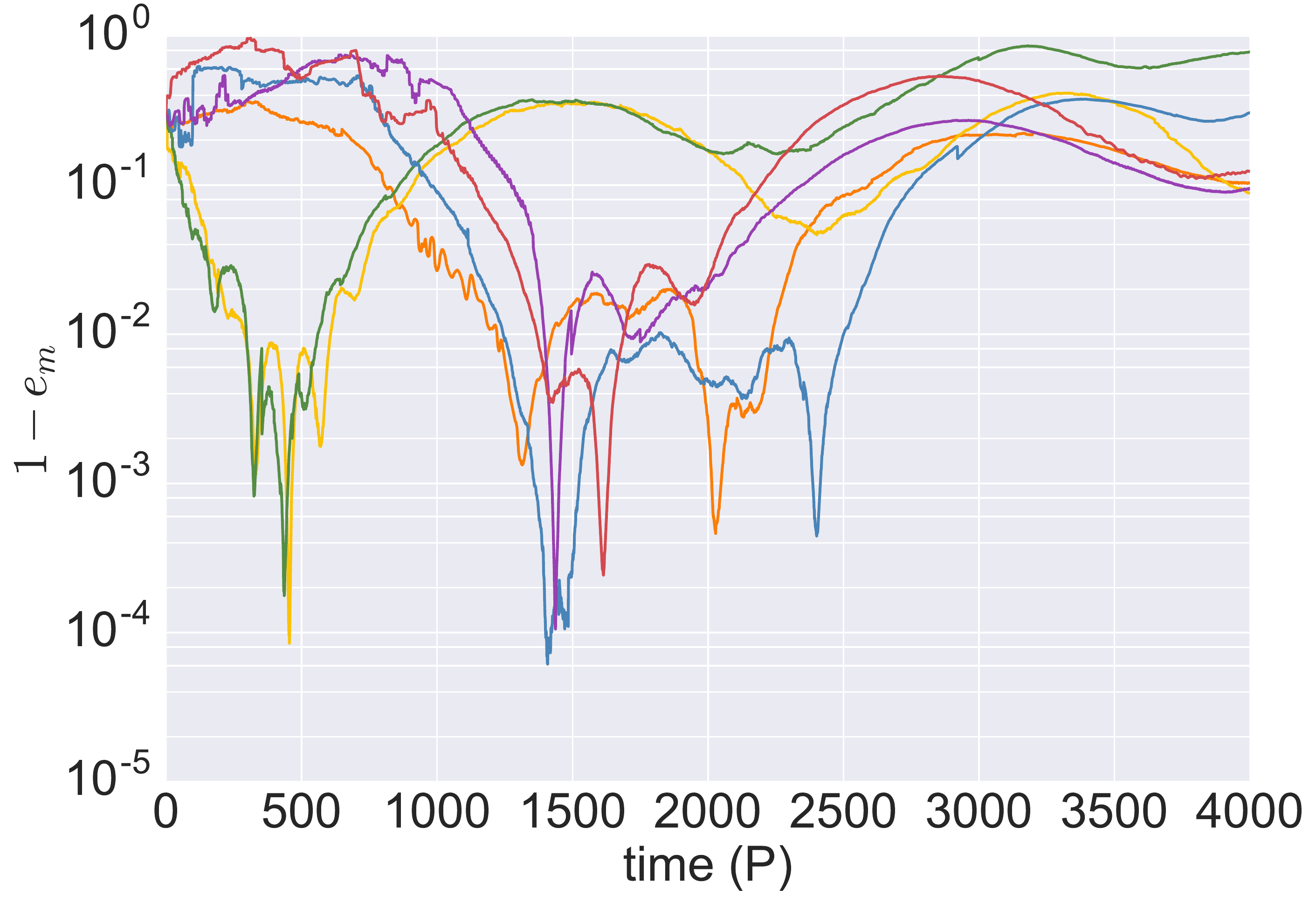}
\caption{\textbf{Eccentricity evolution of small mass bodies} 
Small masses undergo eccentricity oscillations, similar to those of the massive body but in opposite direction and of higher amplitude.  As eccentricities grow the small bodies are in danger of colliding with the central object at perihelion.}
\label{fig:comets}
\end{figure}

The angular momentum exchange between large and small mass bodies in our simulations greatly excites the eccentricities of the small masses. Figure \ref{fig:comets} illustrates the eccentricity evolution of a handful of small masses from the simulations described in section \ref{sec:results} where bodies are initialized with eccentricity $e=0.70$ and $a=[0.9,1.1]$.  As the more massive body is torqued to low eccentricities, the reverse happens to nearby smaller mass bodies.  As a result, some are kicked to very high orbital eccentricities, $e \approx 0.999$. In a real planetary system, they would be in danger of colliding with the central star or inner planets. 

Some of these bodies in our simulations also evolve onto retrograde orbits.  This occurs when they are torqued to low enough orbital angular momenta that they undergo "inclination flips" \citep{2014ApJ...785..116L,Madigan2018}.  This is interesting given  the existence of retrograde orbiters in the outer Solar System such as Drac \citep{2009ApJ...697L..91G} and Niku \citep{2016ApJ...827L..24C}.

To determine whether this mechanism could produce retrograde orbiters in the real Solar System, we use the language of loss cone dynamics \citep{Fra76,Hills81}. The loss cone is the region of phase space in which bodies are in danger of being disrupted by, or colliding with, the central body. 
In angular momentum space, a body is in danger if it is below the loss cone threshold. The loss cone angular momentum threshold is defined as
\begin{equation}
\label{eq:losscone}
J_{LC} \approx \sqrt{2GM_{*}R_{\rm LC}}.
\end{equation}
where $G$ is the gravitational constant and $R_{\rm LC}$ is the radius around the central body where collision/disruption occurs.

The system can be classified into one of two categories, the full or empty loss cone regime, which is determined by the parameter $q$, 
\begin{equation}
\label{eq:peri_losscone}
q = \left(\frac{\Delta J_{p}}{J_{LC}}\right)^2 
\end{equation}
\citep{Lightman&Shapiro1977}.  Again, $\Delta J_p$ is the change in angular momentum of an orbit per orbital period (see equation \ref{eq:jp}). In the full loss cone regime ($q \gg 1$), the average change in angular momentum of a body in the system is large enough that it can jump in and out of the loss cone in one orbital period. In this regime, there are always bodies in the loss cone because the system's dynamics is constantly replenishing its population. In the empty loss cone regime ($q \ll 1$), the average change in angular momentum of a body per orbital period is much smaller than the size of the loss cone. Thus, the system cannot replenish the loss cone population as quickly as bodies in that population are being destroyed. Therefore, the loss cone phase space region will be empty.

Could scattered disk objects evolve onto retrograde orbits via this mechanism in the real Solar System? Previously, we defined $\Delta J_p$ for the massive body (equation \ref{eq:jp}). Here, we are interested in the small mass trans-Neptunian objects. The change in angular momentum per orbital period for these small bodies is
\begin{equation}
\label{eq:mjp}
\Delta J_{p_m} = 2\pi~\beta(e)~\frac{M}{M_{\odot}}~J_c,
\end{equation}
and $q$ is
\begin{equation}
\label{eq:qm}
q = \frac{\pi^2}{8}~e_{M_0}^2~\frac{M^2}{M_{\odot}^2}~\frac{a}{R_\odot}
\end{equation}
assuming that $M_{*} = M_{\odot}$ and $R_{LC} = R_{\odot}$.  To determine whether or not retrograde bodies will be produced in the real Solar System we set $q = 1$ and solve for the mass $M$ needed for the small mass bodies to live in the full loss cone regime.  Assuming this massive body lies at $a = 100$ AU and has an eccentricity $e_{M_0} = 0.7$, we find $M \approx 8.8 \times 10^{-3} ~ M_{\odot}$. We conclude that scattered disk objects clearly exist in the empty loss cone regime and that this new secular mechanism cannot create retrograde orbiters in the Solar System. 
\section{Relevance of mechanism to Real Astrophysical Systems}
\label{sec:relevance}

Here, we explore if this mechanism can detach bodies in the scattered disk, reduce the eccentricity of a hypothetical Planet 9, or be responsible for the pollution of white dwarf surfaces. 

\subsection{How (not) to make Detached Objects}
\label{subsec:detached}
Could this new secular mechanism be responsible for the creation of detached objects in the scattered disk?  One interesting piece of tentative evidence for this is that TNOs with larger diameters seem to have larger perihelion distances and lower eccentricities \citep{2012A&A...541A..92S}.  Although \citet{2012A&A...541A..92S} state that this is most likely an observational bias, we investigate the possibility that the observed correlation is a dynamical artifact.  Our mechanism could provide an explanation for this; high mass bodies (with larger diameters) would have slower initial apsidal precession rates than the rest of the disk and will be subsequently torqued to lower eccentricities via more frequent disk interactions.  We investigate this further and take Sedna \citep{0004-637X-617-1-645} as our sample case, with a perihelion distance of $76$ AU, a semi-major axis of $506$ AU, an eccentricity of $0.85$, an orbital period of $11,400$ years, and an estimated mass of $ \sim 10^{-9} M_\odot$ \citep{SEDNA}.

\begin{figure}[t]
\includegraphics[width=0.45\textwidth]{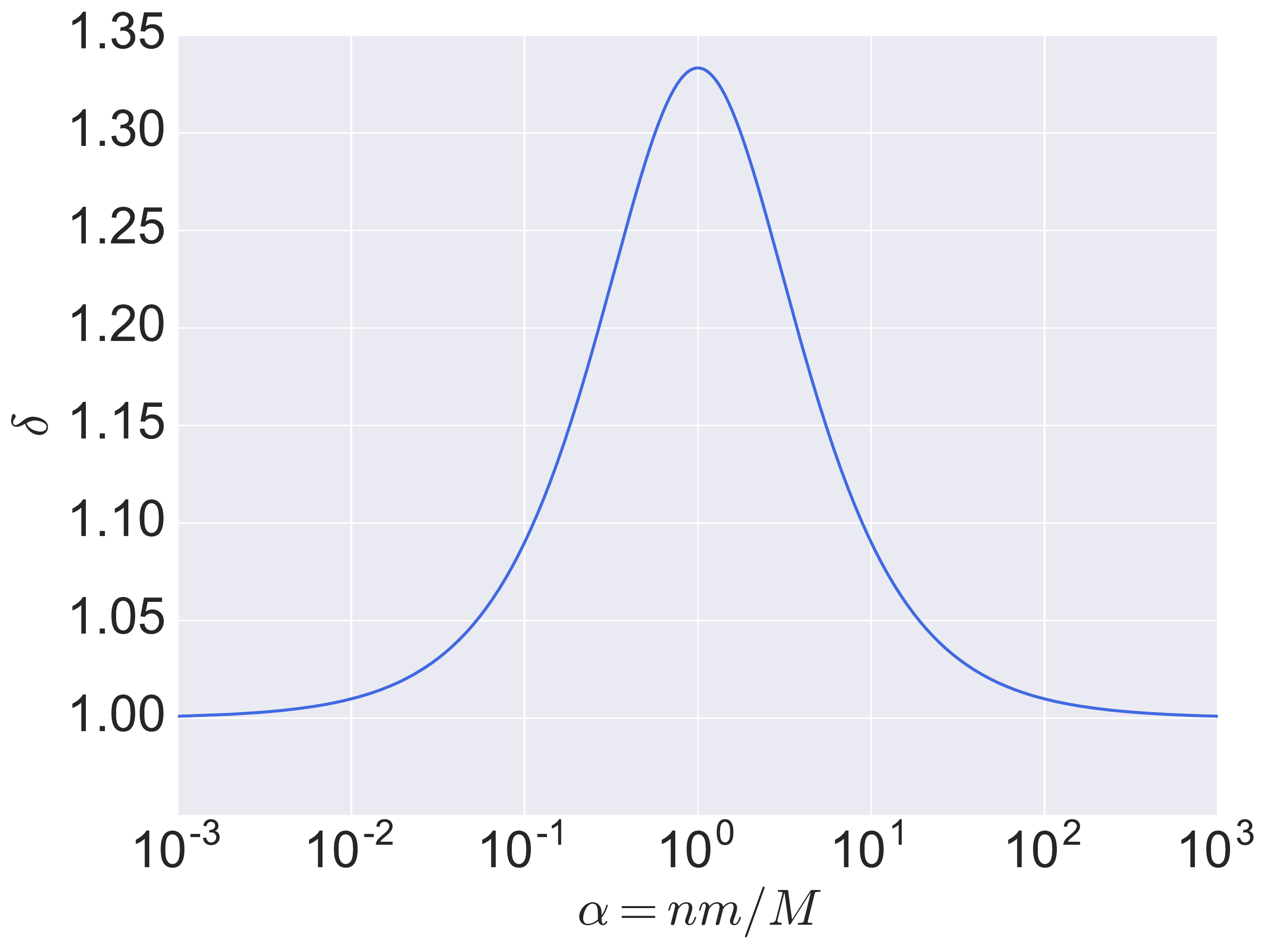}
\caption{\textbf{$\delta$ vs. disk Mass} The interaction term $\delta$ approaches unity for sufficiently high and low disk masses.  The maximum interaction (or transfer of angular momentum) occurs when $nm \sim M$.  The lowest equilibrium eccentricity a massive body is able to attain is determined by maximizing $\delta$.}
\label{fig:delta}
\end{figure}

From equations \ref{eq:e_eq} \& \ref{eq:delta}, the equilibrium eccentricity is: $ 1 - e_{\rm eq_M}^2 = \delta ~ \sqrt{ 1 - e_{\rm m_0}^2}$ where $\delta = (1 + \alpha)^2 / [\alpha^2 + (1 + \alpha)]$.  
In Figure \ref{fig:delta}, we see that the lowest equilibrium eccentricity is obtained when the interaction term $\delta$ is maximized, or when $nm \sim M$.  
Figure \ref{fig:delta} also shows that if $M \gg nm$, $\delta$ asymptotically approaches unity and $e_{eq_M} \approx e_{M_0}  = e_{m_0}$, because the massive body can only take so much angular momentum from the surrounding orbits.  
The same occurs when $M \ll nm$.  
This is because as the disk becomes increasingly massive, the precession rates of all of the scattered disk objects are essentially equivalent and again, $e_{eq_M} \approx e_{M_0}$.  
Therefore, in order for Sedna to achieve a low current eccentricity through this mechanism it must interact with a disk similar to its own mass (i.e. $nm \sim M$). 
Given its current semi-major axis and an assumed starting eccentricity of $e_{M_0} = 0.94$ (corresponding to $p = 30$ AU), its lowest possible \textit{equilibrium} eccentricity is $e_{eq_M} = 0.89$.\footnote{This is not the lowest \textit{eccentricity} Sedna can achieve. 
Its equilibrium eccentricity corresponds to a specific angular momentum of $J_{{eq}_M} = J_c ~ (1-e_{{eq}_M}^2)^{1/2}$.  
The amplitude of the angular momentum oscillations are $\delta J = J_{{eq}_M} - J_{M_0}$ where $J_{M_0} = J_c ~ (1-e_{M_0}^2)^{1/2}$.  
The highest possible angular momentum it can reach from disk interactions is $J_{M_h} = J_{M_0} + 2\delta J$.  
Therefore, the lowest eccentricity Sedna can achieve via this mechanism is $e_{M_l} = \sqrt{1-\left({J_{M_h}}/{J_c}\right)^2} \approx 0.82$.}  
From equation \ref{eq:p-osc}, we calculate Sedna's oscillation timescale of $t = 1.2 \times 10^{13}$ years!  This clearly shows that Sedna could not have undergone detachment via this mechanism.

We conclude that, although our simulations demonstrate the robustness of our mechanism in increasingly realistic conditions, it is not relevant in the mass regime of the scattered disk on solar system timescales (despite that it is valid in the scattered disk's orbital regime).  We chose a very high mass for the massive body in our simulations, roughly equivalent to a Jupiter mass, with a comparably high disk mass orbiting a solar mass star. Since $t_{\rm osc} \propto \frac{\Mbh}{nm} ~ P$ (equation~\ref{eq:p-osc}), detachment of massive bodies in our simulations occurs within hundreds of orbital periods.  Sedna and the real scattered disk population have much lower masses.  As a result, it is not be possible for minor planets to detach from the scattered disk via this mechanism.

This is important for the inclination distributions of different mass populations as described in section \ref{sec:results}.  After the inclination instability has raised the average inclination of the small mass' orbits to $\sim 30\degree$, massive bodies have on average lower inclinations than their less massive counterparts.  This is directly due to their secular exchange of angular momentum with small mass population. Since massive bodies initially drop to lower eccentricities, their orbits become harder to torque off the orbital plane.  The extremely long timescale for this means it should not noticeably affect the observed inclination distributions of minor planets in the outer Solar System.

In general, disks of very high eccentricity orbits will struggle to appreciably lower a massive body's eccentricity via this mechanism.  In a very high eccentricity disk ($e \gtrsim 0.9$), all orbits have low specific orbital angular momentum to begin with.  Therefore, small masses can only donate a small amount of angular momentum to their more massive counterparts.  As a result, a large mass will not significantly lower its eccentricity even if $\delta$ is maximized.  This becomes important when considering circularizing the orbits of bodies like the hypothetical Planet 9 \citep{2016AJ....151...22B}, which could be embedded in a disk of high eccentricity planetesimals. 

\subsection{Circularizing Planet 9?}
\label{subsec:Planet9}
The hypothetical Planet 9 has a predicted mass of $\sim 5-10 M_{\oplus}$ and a semi-major axis of $\sim 700$ AU \citep{2016AJ....151...22B,2017AJ....154..229B}. It is not obvious how such a large planet could have made its way into the outer Solar System.  It could not have formed in-situ \citep{2016AJ....151...22B} and is unlikely to have been captured in the sun's natal star forming region \citep{2017MNRAS.472L..75P}.  However, \citet{1999Natur.402..635T} \& \citet{2007Icar..189..196L} demonstrate that massive protoplanets, forming alongside the ice giants in the early Solar System, can be scattered outward onto high eccentricity orbits while the young giants are clearing debris from their orbital domains. \citet{2016ApJ...823L...3L} show that the probability of this occurring $\lesssim 5 \%$. 
If Planet 9 exists and was formed by such processes, it would be a challenge to explain its currently low(er) eccentricity; which is predicted to be $\sim 0.5-0.8$ \citep{2017AJ....154..229B}.  This is because scattering Planet 9 from the inner Solar System onto a high semi-major axis orbit would inevitably leave it with a perihelion still in the region of the giant planets. 
Could the hypothetical Planet 9 have been scattered outward in the distant past and then decreased its orbital eccentricity via our detachment mechanism?  As before, we maximize the interaction term $\delta$ and minimize Planet 9's equilibrium eccentricity by assuming that it interacts with a disk equivalent to its mass.  If we assume it was scattered from $\sim 10$ AU, it would have an initial eccentricity of $e_{M_0} = 0.99$. It's minimum equilibrium eccentricity would then  be $e_{eq_M} = 0.97$.  This value is far higher than Planet 9's predicted eccentricity which tells us that Planet 9 could not have reached its current eccentricity via this mechanism. 
 
\subsection{Where could this mechanism be important?}
Although this mechanism does not seem important in the outer Solar System, it could be for other Keplerian systems which host disks of sufficiently high mass (i.e. $\Md \sim 10^{-3} \Mbh$).
For example, disks of stars on eccentric orbits around super massive black holes (SMBHs) could preferentially circularize the most massive bodies while feeding the central SMBH. 
Additionally, 20\% of hydrogen-rich white dwarfs have surfaces polluted heavily by metals \citep{1980ApJ...242..195C,2008ApJ...685L.133M}.  This is odd considering that the sinking time for metals on their surfaces is quite short, indicating that metals are being continuously supplied to their surfaces on rapid timescales.  It is unclear how white dwarfs could continually accumulate metals on their surfaces, but one suggestion is a continuous bombardment of asteroids from a planetesimal disk \citep{1986ApJ...302..462A,0004-637X-572-1-556,2012ApJ...747..148D}.  A massive planet ($\sim M_{\rm Jup}$) embedded in a planetesimal disk roughly equal to its mass could undergo eccentricity oscillations via this mechanism and, in the process, generate a flux of bodies that reach high enough eccentricities such that they interact with the central white dwarf.  If we assume that these planetesimal disks are similar to the ones we simulate in section \ref{sec:results} (i.e. the disk orbits have starting eccentricities $e_{m_0} \approx 0.7$ and the planet has an equilibrium eccentricity of $e_{eq_M} \approx 0.55$; which corresponds to our numerical results), we can use equation \ref{eq:p-osc} to find that the bombardment timescale for a solar mass white dwarf is $\sim 1.5 \times 10^3$ orbital periods. If the planet has a semi-major axis of 10 AU the bombardment timescale would be $\sim 5 \times 10^4$ years.

\section{Summary and Discussion}
\label{sec:disc}

In this paper we show that within an axisymmetric disk of eccentric orbits, massive bodies exchange orbital angular momentum with less massive bodies in a process involving differential apsidal precession and secular gravitational torques. As a result, massive bodies undergo oscillatory behavior in orbital eccentricity (and perihelion distance), with an equilibrium value lower (higher) than their starting values. 

Using $N$-body simulations of increasing realism, we find that this mechanism is robust in outer Solar System conditions.  These conditions are, namely, eccentricities and semi-major axes consistent with the scattered disk population, scattering forces from the giant planets, and a distribution of masses that follow a $m^{-5/3}$ power law.  Unfortunately, we find that this mechanism is not capable of detaching minor planets from the scattered disk within the age of the Solar System.  However, this mechanism could be relevant for the pollution of white dwarf surfaces and disks of stars around supermassive black holes.

\acknowledgments
We gratefully acknowledge support from NASA Solar System Workings under grant 80NSSC17K0720. Simulations in this paper made use of the {\tt REBOUND} code which can be downloaded freely at https://github.com/hannorein/rebound. We thank Dan Tamayo and Hanno Rein for their infinite patience. 
This work utilized the RMACC Summit supercomputer, which is supported by the National Science Foundation (awards ACI-1532235 and ACI-1532236), the University of Colorado Boulder, and Colorado State University. The Summit supercomputer is a joint effort of the University of Colorado Boulder and Colorado State University.
We additionally thank Konstantin Batygin for the discussion that motivated section \ref{subsec:Planet9} of this paper.
Lastly, we thank Dr. Richard Fienberg and the American Astronomical Society (AAS) for giving us the opportunity to discuss our early results at the \nth{232} AAS Conference in Denver, Colorado.
\software{REBOUND (Rein and Liu 2012, Rein and Spiegel 2015)}

\bibliography{MasterRefs}
\end{document}